# Detector with a profile-based cathode and a two-coordinate pad-strip readout system

*N. A. Kuchinskiy[1*], V. A. Baturitskii, N. P. Kravchuk[1], A. S. Korenchenko[1], N. V. Khomutov[1],*

*V. S. Smirnov[1], V. A. Chekhovskii[2], S. A. Movchan[1], F. E. Zyazyulya[2]*

[1] – Joint Institute for Nuclear Research, 141980 Dubna, Moscow region, Russia

[2] – National Scientific and Educational Centre of Particle and High Energy Physics of the Belarussian State University, M. Bogdanovich street 153, 220040 Minsk, Belarus

[*]corresponding author, tel. +7-496-21-65173, kuchinski@jinr.ru

**A detector with a profile-based cathode and a pad-strip cathode readout system is experimentally investigated. Cathode pads arranged along each anode wire are diagonally interconnected and form strips that cross the detector at an angle with respect to the anode wire. Two coordinates from the cathodes and one from the anode wire allow identification of tracks in high multiplicity events with a single detector plane.**

Track detectors with a profile-based cathode, the best known of which are Iarocci tubes [1, 2] have been widely used in high energy physics now. Their major advantage is a simple design and possible modular structure, i.e. a possibility of assembling large-area detectors from standard elements. Reliability of these detectors is based on the fact that malfunction of one wire does not cause a failure of the entire detector.

A possibility of determining coordinates along the wire with the aid of outer strips perpendicular to the anode wires in Iarocci tubes was investigated in many papers [3,4]. This kind of readout provides spatial resolution along the wire of about 300-400 μm but does not solve the problem of multiple track registration in a single detector plane.

We investigated the possibility of determining the coordinate along anode wire of the profile-based detector by using a two-coordinate system of pads connected into strips at symmetric angles relative to the anode wire (Fig. 1). So the signals from a fired anode wire are induced only on the pads belonging to the same cell as the wire due to the shielding effect of the rectangular cell side walls.

Thus, using two coordinates from the cathode strips and the third coordinate obtained from the anode wire it becomes possible to resolve particle tracks in high multiplicity environment in one detector plane.

The experimental detector was made on the basis of the standard Iarocci tube developed for the COMPASS experiment: an open aluminum profile with 0.6 mm thick walls. The inner size of eight cells is 9.4 mm × 9.4 mm. The distance between the cell centers is 10 mm. The anodes are made of gold-plated tungsten wire 30 μm in diameter.

The board with the pad-strip cathode system is an 80 mm × 40 mm plate made of glass fiber epoxy laminate covered on one (outer) side with copper foil which is etched to form the pads. The pads are interconnected on the opposite side of the board.

The pads used in these measurements are 9 mm × 5 mm (9 mm across the anode wire and 5 mm along it). The pad spacing is 5 mm, and the anode wire runs exactly along the pad axis. High voltage is applied to the anode wires and signals are picked up from them through high-voltage capacitors. The precise mounting of the cathode pads relative to the anode wire was ensured by recesses in the board into which the aluminum profile edges fit.

The detector layout is shown in Figs. 1 and 2. The detector with the pad-strip cathode was placed between a collimated $^{90}$Sr source and a 5 mm × 5 mm scintillation counter. A 0.5 mm wide slit was made in the aluminum profile along the anode wire to minimize multiple scattering of the source particles. A similar slit was made in the pad board on the outer case.

The measurements were triggered by coincidence of the signals from the scintillation counter and an anode wire. The detector itself could be moved relative to the source and the scintillator using a precise mechanical positioning system with an accuracy of 10 μm.

Pad-strip signals were amplified by the 16-channel cathode amplifier based on the KATOD–1 integrated circuit [5] and fed into VME CAEN V1720 waveform digitizer (12-bit, 250 MHz) [6]. The anode wire signals were amplified by the AMPL-8.3 eight-channel amplifier [7] used also in the experiments D0 (FNAL) [8] and COMPASS (CERN) [9]. Both integrated circuits of the amplifier were developed at the National Center of Particles and High-Energy Physics in Minsk and its parameters are presented in Table 1.

The coordinate along the wire is determined by finding the center of gravity of the charge distribution on the strips. It is noteworthy that the waveform digitizing technique provides extra possibilities. Good temporal resolution of the electronics allows the drift time and accordingly the coordinate perpendicular to the wire to be measured using the wire signals.

Figure 3 shows linearity of the measured coordinate (Y) in respect to the $^{90}$Sr source position (X). By summing the charges induced on the pad-strips, their correlation with the charge on the anode is measured (Fig. 4). The total charge induced on the pads is directly proportional to the anode charge and amounts to about 25% of the total charge on the anode.

The pad-strip charge distribution ( in relative units) is presented on Fig.5, left. A profile of the collimated $^{90}$Sr source, 2 mm diameter, at the position of X=28 mm along the anode wire is shown on Fig.5, right. The RMS of the distribution is about 1 mm. The results were obtained at the high voltage of 3.40 kV.

**REFERENCES**

[1] R. Appel et al. A Large Acceptance, High Resolution Detector for Rare $K^+$-Decay Experiments, Nucl. Instr.&Meth, A479, 2002, 349-406.

[2] E. Iarocci. Plastic streamer tubes and their applications in high energy physics, Nucl. Instr.&Meth. A217, 1, 30-42.

[3] N. Khovansky, V. Malyshev, V. Tokmenin et al. Spatial resolution of profile-based detectors with external pick-up strips, Nucl. Instr.&Meth, A351, 1994, 317-329.

[4] G. Bauer, A. Bettini, G. Busetto et al. Resolution of plastic streamer tubes with analog readout, Nucl. Instr.&Meth, A260, 1987, 101-113.

[5] I. A. Golutvin et al., The Catod-1 strip Readout ASIC for the Cathode Strip Chamber, JINR, E13-2001-151, 2001.

[6] CAEN Electronic Instrumentation, 2010 Products Catalog, V1720.

[7] G. D. Alexeev et al. The eight-channel ASIC bipolar transresistance amplifier D0M Aml-8.3. Nucl. Instr.&Meth, A462, 2001, 494-505.

[8] V. M. Abazov et al., The Upgraded D0 Detector, Nucl. Instr.&Meth. A565, 2006, 463.

[9] The COMPASS Experiment at CERN, Nucl. Instr.&Meth., A577, 2007, 455-518.

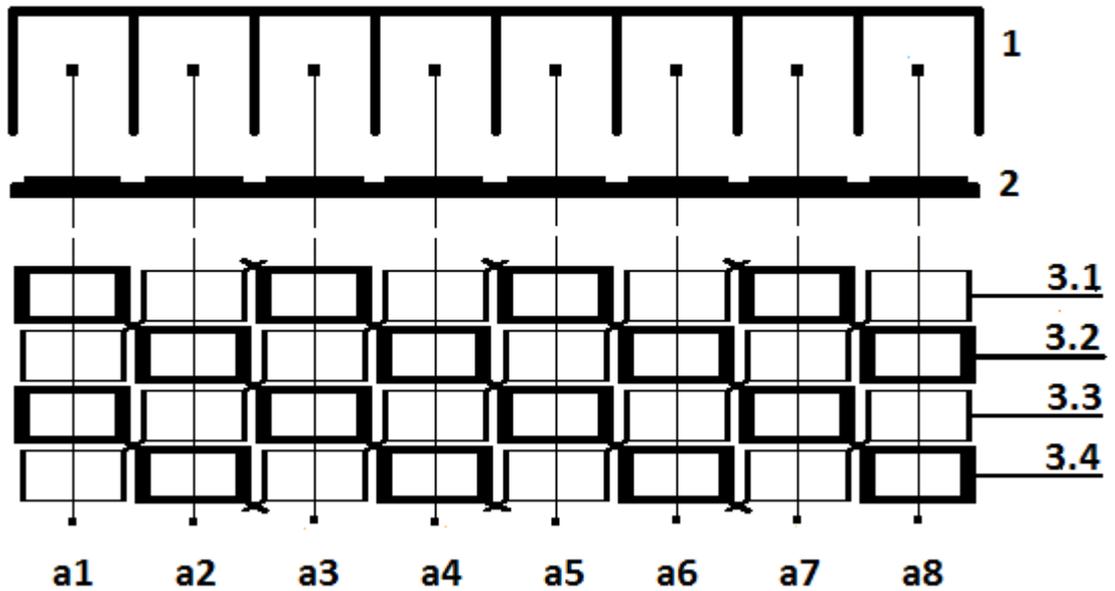

Fig. 1. Sketch of Iarocci 8-channel module (1) with pad-strip read-out board (2); a1-a8 – wires read-out and 3.1-3.4 – strips read-out.

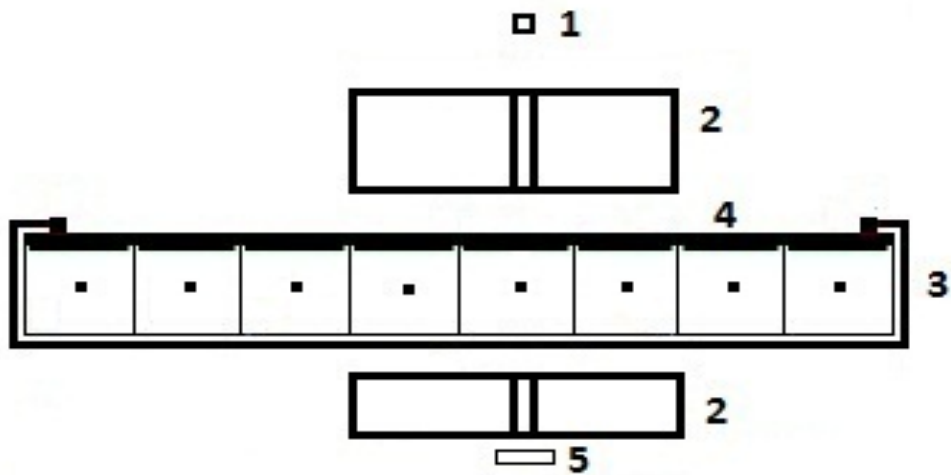

Fig. 2. Experimental layout. 1 – $^{90}$Sr source; 2 – collimator, 2 mm diameter; 3 – Iarocci 8-channel module; 4 – pad-strip read-out board; 5 – scintillator 5 mm× 5 mm.

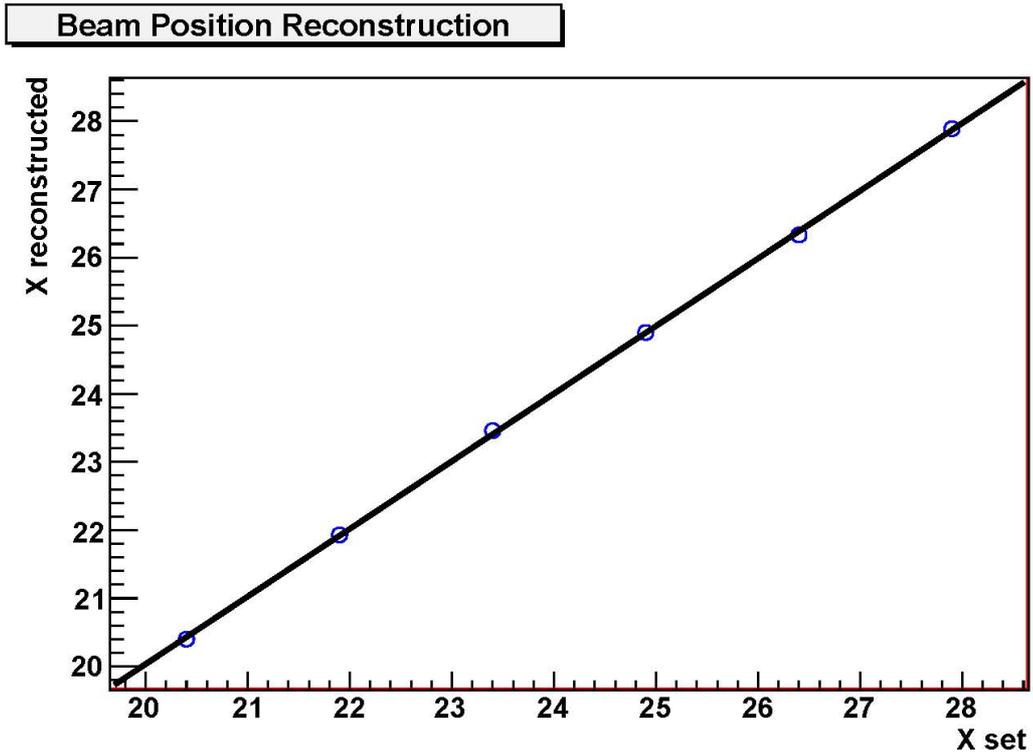

Fig. 3. Linearity of measured coordinate Y in respect to $^{90}$Sr source position X in mm.

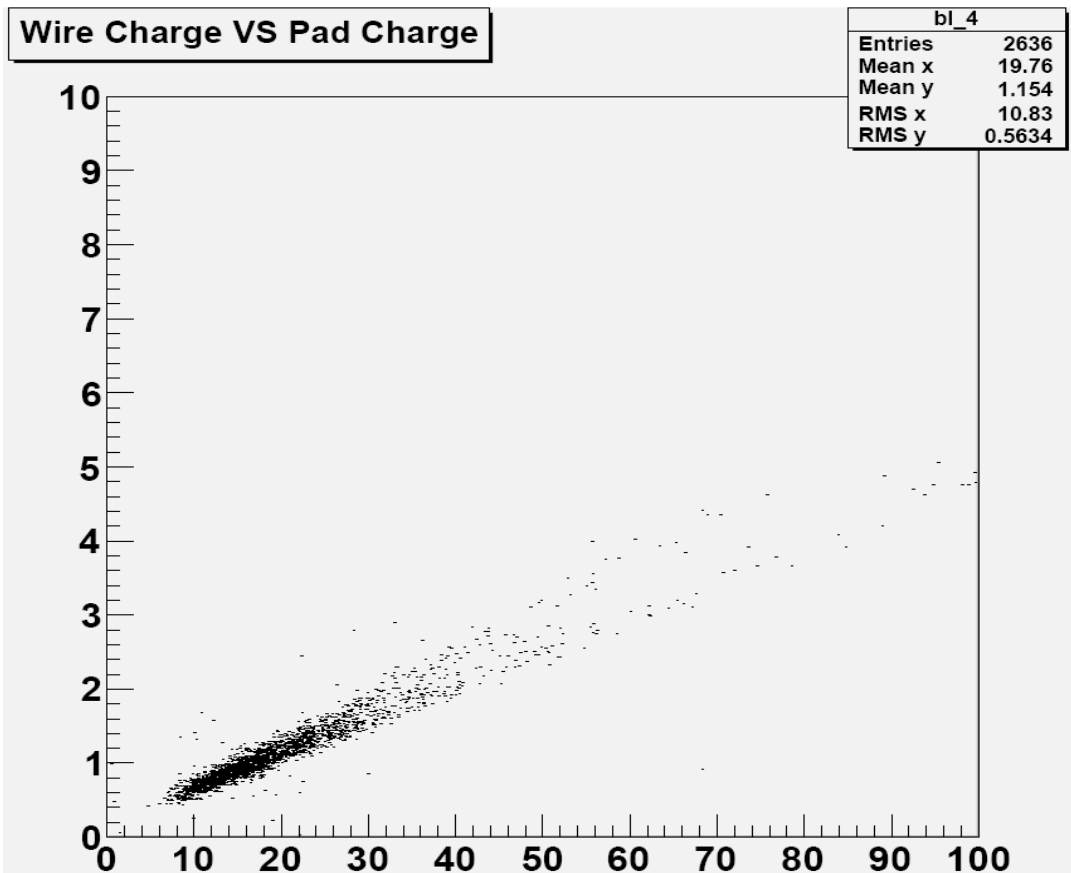

Fig. 4. Correlation between anode wire signals (Y) and the sum of the charges (X) induced on the pads at HV=3.40 kV. Horizontal and vertical scales – relative units.

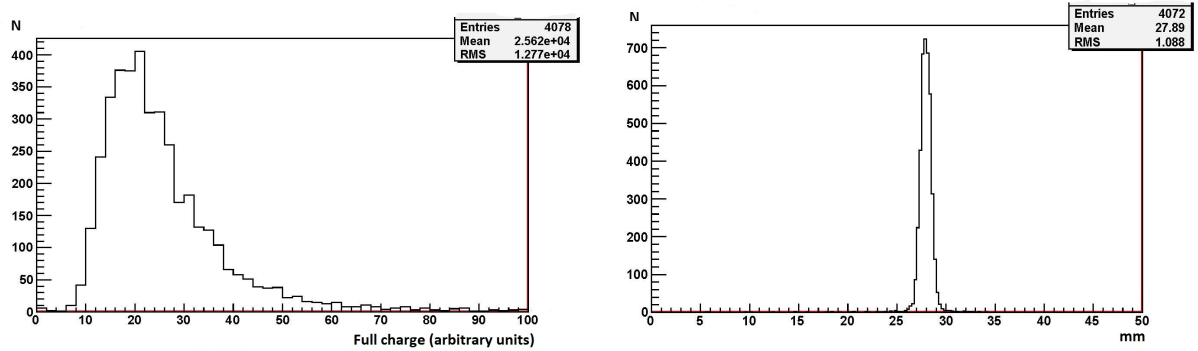

Fig. 5. Pad-strip charge distribution (relative units) – left. $^{90}$Sr collimated source (2mm diameter) profile at the position of X=28 mm – right. HV=3.40 kV.

Table 1. Parameters of front-end electronics used in the measurements

| Parameter | Ampl-8.3 | Katod-1 |
|---|---|---|
| Input resistance, $\Omega$ | 50 | 50 |
| Conversion factor | 70 mV/µA | < 6 mV/fC |
| Scattered power, mV/channel | < 80 | < 17 |
| Equivalent input noise current (RMS), nA | 40 ($C_{detector}$ = 0)<br><br>60 ($C_{detector}$ = 60 pF) | |
| Equivalent input noise charge (RMS), electrons | | 2400 + 12/pF |
| Number of channels in package | 8 | 16 |